\documentclass[aps,prl,twocolumn,showpacs,groupedaddress]{revtex4}  
\usepackage{graphicx}  
\usepackage{dcolumn}   
\usepackage{bm}        
\usepackage{amssymb}   
\usepackage{epstopdf}

\begin{document}
\def\d0{D\O}
\newcommand{\bcdec}{${B}_{c}^{\pm}  \rightarrow J/\psi\mu^{\pm} + X\:$}
\newcommand{\bc}{$B_c^{\pm} \:$}
\newcommand{\jpsi}{$J/\psi~$}
\newcommand{\Jpsi}{J/\psi}
\hyphenation{EVTGEN}


\hspace{5.2in} \mbox{FERMILAB-PUB-08-136-E}

\title{Measurement of the lifetime of the \boldmath $B^{\pm}_c$ meson
in the semileptonic decay channel}
%
\author{V.M.~Abazov$^{36}$}
\author{B.~Abbott$^{75}$}
\author{M.~Abolins$^{65}$}
\author{B.S.~Acharya$^{29}$}
\author{M.~Adams$^{51}$}
\author{T.~Adams$^{49}$}
\author{E.~Aguilo$^{6}$}
\author{S.H.~Ahn$^{31}$}
\author{M.~Ahsan$^{59}$}
\author{G.D.~Alexeev$^{36}$}
% ...

\affiliation{\vspace{0.1 in}(The D\O\ Collaboration)\vspace{0.1 in}}
% ...
\affiliation{$^{6}$University of Alberta, Edmonton, Alberta, Canada,
                Simon Fraser University, Burnaby, British Columbia, Canada,
                York University, Toronto, Ontario, Canada, and
                McGill University, Montreal, Quebec, Canada}
% ...
\affiliation{$^{29}$Tata Institute of Fundamental Research, Mumbai, India}
% ...
\affiliation{$^{31}$Korea Detector Laboratory, Korea University, Seoul, Korea}
% ...
\affiliation{$^{36}$Joint Institute for Nuclear Research, Dubna, Russia}
% ...
\affiliation{$^{49}$Florida State University, Tallahassee, Florida 32306, USA}
% ...
\affiliation{$^{51}$University of Illinois at Chicago,
                Chicago, Illinois 60607, USA}
% ...
\affiliation{$^{59}$Kansas State University, Manhattan, Kansas 66506, USA}
% ...
\affiliation{$^{65}$Michigan State University,
                East Lansing, Michigan 48824, USA}
% ...
\affiliation{$^{75}$University of Oklahoma, Norman, Oklahoma 73019, USA}
% ...
\date{May 16, 2008}

\begin{abstract}
Using approximately 1.3 fb$^{-1}$ of data collected by the D0 detector between 2002 and 2006, we measure the lifetime of the $B_c^{\pm}$ meson in the \bcdec final state. A simultaneous unbinned likelihood fit to the $J/\psi + \mu$ invariant mass and lifetime distributions yields a signal of $881 \pm 80\thinspace {\mathrm{(stat)}}$ candidates and a lifetime measurement of $\tau(B_c^{\pm}) = 0.448^{+0.038}_{-0.036} \thinspace {\mathrm{(stat)}} \pm 0.032 \thinspace {\mathrm{(syst)}}\thinspace {\mathrm{ps}}.
$
\end{abstract}
\maketitle 


One of the most interesting mesons that can be studied at the Tevatron is
the $B_c^{\pm}$. Unlike most $b$ hadrons, the $B_c^{\pm}$ meson
comprises two heavy quarks ($b$ and $c$) that can each decay with significant contribution to the decay rate, or they can participate in an annihilation mode.
The $B_c^{\pm}$ meson is therefore predicted~\cite{bc_pred,bc_pred2} to have a lifetime of
only one-third that of the other $B$ mesons, the shortest
of all weakly-decaying $b$ hadrons.
Example final states where the $c$ quark acts as a spectator while the $b$ quark decays
weakly are
$B_c^{\pm}\rightarrow \Jpsi \pi^{\pm}$, $B_c^{\pm}\rightarrow \Jpsi D_{s}^{\pm}$,
and $B_c^{\pm}\rightarrow \Jpsi \ell^{\pm}\nu$.



In this Letter we present a measurement of the lifetime of the $B_c^{\pm}$ meson in the \bcdec final state with $J/\psi \rightarrow \mu^{+}\mu^{-}$, using approximately 1.3 fb$^{-1}$~\cite{lumi} of data collected with the D0 detector~\cite{d0det} at the Fermilab Tevatron collider. The detector components most important to this analysis are the central fiber tracker (CFT), the silicon microstrip tracker (SMT), and the muon system~\cite{run2muon}.
An inclusive muon triggered sample is used where events selected only by lifetime-biasing triggers are excluded.
The invariant mass of the resulting trimuon system is used to help separate
the signal and background components and determine their normalizations.  

The decay length used to extract 
the $B_c^{\pm}$ lifetime is measured as the distance between the 
reconstructed primary proton-antiproton interaction vertex and the secondary 
vertex formed by the \jpsi and the third muon.
The presence and behavior of the 
$B_c^{\pm}$ signal is demonstrated using mass fits
following decay length requirements. 
We construct models of the lifetime distributions of signal
and various background components and then perform a simultaneous fit
to the trimuon invariant mass and lifetime distributions 
to measure the lifetime of the $B_c^{\pm}$ meson.


To simulate $B_c^{\pm}$ properties in this final state and to determine appropriate 
selection criteria, Monte Carlo (MC) signal samples of
$B_c^{\pm} \rightarrow J/\psi(\rightarrow\mu^+\mu^-) \mu^{\pm} \nu$
are generated using the standard D0 simulation chain, including the {\sc pythia} event generator~\cite{pythia} interfaced with the {\sc evtgen} decay package~\cite{evtgen} followed by full {\sc geant}~\cite{geant}  modeling of the detector response and event reconstruction.
For the simulated signal samples, the ISGW
semileptonic decay model~\cite{isgw} for $B_c^{\pm}$ is used. A separate sample using a phase-space decay model is generated for systematic studies.  
Another possible decay of the $B_c^{\pm}$ is $B_c^{\pm}\rightarrow \psi(2S)\mu^{\pm} + X$ where $\psi(2S)\rightarrow J/\psi \pi^{+} \pi^{-}$, and a sample of this mode is 
generated as well. 
To model one of the backgrounds, a large MC sample of inclusive $J/\psi$ events, including $b$ production
via gluon splitting and flavor excitation,
is used, only requiring a generator-level 
$J/\psi \rightarrow \mu^+\mu^-$ decay.


We begin by selecting a subsample of events
containing at least one $J/\psi \rightarrow \mu^+\mu^-$ candidate
with at least two muons of opposite charge
reconstructed in the CFT, SMT, and the muon system.
The track of each muon must have transverse momentum $ p_T> 1.5$~GeV, and match hits
in the muon system,
or it must have $p_T > 1.0$~GeV and a calorimeter
energy deposit consistent with that of a minimum-ionizing particle. 
For at least one of the muons, hits are required in all 
three layers of the muon detector, and each must have at least
two hits in the CFT and at least one hit in the SMT.
The signal region is defined in terms of the dimuon mass to be
$2.90 < M(\mu^+\mu^-) < 3.26$~GeV.
The muon momenta are adjusted according to
a mass-constrained fit to the known $J/\psi$ mass~\cite{PDG}.

Once a \jpsi is found, 
an additional third track that can be associated with the \jpsi vertex is sought.
The following cuts are applied to the resulting $J/\psi$+track candidate: 
the third track must have at least two hits in the SMT, the extrapolation of the three-track momentum must be consistent with coming from the primary vertex, 
$p_T({\mathrm{third \thinspace track}}) > 3$~GeV, 
$p({\mathrm{third \thinspace track}}) > 4$~GeV, 
$p_T(J/\psi + {\mathrm{track}}) > 5$~GeV, 
the $\chi^2$ probability to form a common vertex is greater than 1\%, 
angle between the  \jpsi  and third track~$<$~1 rad, and 
$\cos\theta < 0.99$ where $\theta$ is the three-dimensional angle between any two muons. 
If more than one $J/\psi$+track candidate is present in an event, 
the candidate with the lowest $\chi^{2}$ of the $J/\psi$+track vertex 
is selected.
To be considered a signal candidate, the third track must be identified as a muon: it  must have hits in all three layers of the muon detector and have timing signals in the muon scintillator detectors consistent within 10~ns of the beam
crossing to reduce contamination from cosmic rays. The mass of  the $J/\psi$+$\mu$ candidate is required to be  
in the range $3< M(J/\psi \mu) < 10$~GeV, resulting in 
a sample containing 14753 events. 

The invariant mass of the $J/\psi$+$\mu$ 
can be used to
characterize and separate each of the components that contribute to the $J/\psi$+$\mu$ candidate sample. There are six contributions (one signal, and five backgrounds): $B_c^{\pm}$ signal (\textit{SI});  a real $J/\psi$ associated with a ``fake'' muon due to a track (\textit{JT}); fake \jpsi mesons from combinatorial background (\textit{CB}); a real muon forming a vertex with a real \jpsi where neither is from a $B_c^{\pm}$ decay (\textit{JM}); $B^{+}\rightarrow J/\psi K^{+}$ followed by the decay in flight of  $K^{+} \rightarrow \mu^{+} \nu$;  and a $c\bar{c}$ contribution, where a prompt $J/\psi$ is associated with a muon (\textit{PR}). Each component and the determination
of its mass template is described below.

The signal mass template is determined from the signal MC sample.
Theoretical estimates predict the $B_c^{\pm}\rightarrow J/\psi \mu^{\pm}+X$ branching fraction to be approximately 5 to 100 times larger than that of $B_c^{\pm}\rightarrow \psi(2S)\mu^{\pm}+X$~\cite{bc_pred,theoryfeed}.  This difference gives 13\% to 0\% feed-down contribution and we take (6.5 $\pm$ 6.5)\% in the analysis. 

The invariant mass of the $J/\psi$+track in the data sample is used to model the \textit{JT} component.  The rate of what are denoted fake muons is small and primarily due to decays in flight of $\pi^{\pm} \rightarrow \mu^{\pm} \nu$ and $K^{\pm} \rightarrow \mu^{\pm} \nu$. The $B^{+}\rightarrow J/\psi K^{+}$ decay is used to measure the
contribution of this component.
Fits are made to the $B{^+}$ mass peak in the $J/\psi$+$\mu$ sample 
and the $J/\psi$+track sample, and the ratio of the number of 
$B^{+}$ events in the two samples is taken as the fraction of 
events that are due to a real \jpsi but a fake $\mu$. 
Contributions due to $B^+\rightarrow J/\psi \pi^+$ or $B_c^+ \rightarrow J/\psi \pi^+$ are estimated
to be negligible. 

To describe the \textit{CB} component, a normalized 
mass template is formed from events 
in the \jpsi mass sidebands. 
The sideband regions are defined to be 
events with $M(\mu^+\mu^-)$ in the range  2.62--2.80~GeV or  
3.40--3.58~GeV.
The normalization is taken from the fitted number
of background events in the signal region under the $J/\psi$ mass peak.

The \textit{JM} component represents a significant background that is dominated by $b\bar{b}$ backgrounds, 
where one long-lived $b$ hadron decays to $J/\psi + X$ 
and the other decays semileptonically to a 
muon (or via a cascade decay $b \rightarrow c \rightarrow \mu$).  
The requirement that the $J/\psi$ and $\mu$ be close in angle increases the relative
acceptance for $b\bar{b}$ production via gluon splitting.
To model this background, the \jpsi QCD MC
is used with the requirement that the parent of the \jpsi does 
not arise from a prompt $B_c^{\pm}$ meson, $B^{\pm}$, or  $c\bar{c}$ (the latter two components
are estimated using the data and described below). 

For the $B^+$ component, a fit is made to the mass peak of the $B^{+}$ in the $J/\psi$+$\mu$ data sample. 
This fitted distribution 
is then used as a mass template for the $B^+$ component, thus reducing the uncertainty in the modeling of the width of the mass peak.

Candidates with $L_{xy}<0$, where $L_{xy}$ is the transverse decay length defined as the displacement of the $J/\psi$+$\mu$ vertex from the primary vertex~\cite{pvfind} projected onto the direction of the transverse momentum vector of the $J/\psi$+$\mu$ system, are used to estimate the mass template of the \textit{PR} component.  

\begin{figure}[htb]\begin{center}
\centerline{\includegraphics[width=6cm]{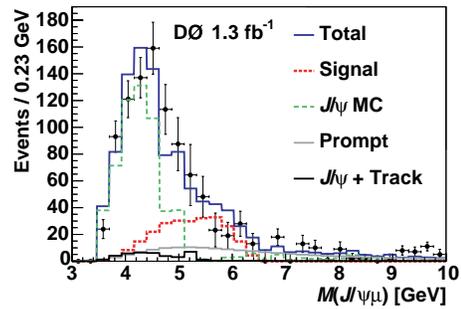}}
\caption{Fit to the mass of the \jpsi + $\mu$ vertex with \jpsi mass sideband and $B^+$ components subtracted and decay length significance $L_{xy}/\sigma(L_{xy})>4$ required. (color online)}
\label{mjpsimufit}
\end{center}
\end{figure}

To check the validity of the modeling of the $M(J/\psi \mu)$ 
distribution, a fit is first made on the mass
distribution of the $J/\psi$+$\mu$ sample using
the templates of the six contributions described above. 
Separate fits are made as the requirement on
$L_{xy}$ is raised to increasingly suppress background.  
Good agreement of the fitted mass components
is observed at all $L_{xy}$ values.  
To  
further check for the presence of the $B_c^{\pm}$ signal,
a requirement is placed on the transverse decay length significance: 
$L_{xy}/\sigma(L_{xy}) > 4$, where 
$\sigma(L_{xy})$ 
is the uncertainty on $L_{xy}$. 
Figure~\ref{mjpsimufit} shows the
fit to the mass distribution after subtracting the \jpsi sideband  and  $B^+$ 
components. In this sample, the statistical significance of the $B_c^{\pm}$ signal component
is $6.4\sigma$.

The lifetime of the $B_c^{\pm}$, $\tau$, is related to the transverse
decay length by $L_{xy} = c \tau \cdot \frac{p_T(B_c^{\pm})}{m(B_c^{\pm})}$, where 
$p_T$ and $m$ are the transverse momentum and rest mass of the $B_c^{\pm}$,
respectively. When the $B_c^{\pm}$ meson decays semileptonically, it cannot be fully
reconstructed due to the escaping neutrino, and thus $p_T(B_c^{\pm})$ cannot be determined.  The $p_T$ of the $J/\psi +\mu$ system is used
instead as an approximation.
A correction factor, $K = p_T(J/\psi\mu)/p_T(B_c^{\pm})$, determined using signal MC,  is
introduced to estimate $p_T(B_c^{\pm})$.
To obtain the $B_c^{\pm}$ lifetime, the visible-proper decay length (VPDL) is measured,
defined as $L_{xy}\frac{m(B_c^{\pm})}{p_T(J/\psi\mu)}=\frac{c{\tau}}{K}.$

The $K$-factor distribution is applied statistically by smearing
the exponential decay distribution when extracting  $c\tau(B_c^{\pm})$
from the VPDL distribution in the lifetime fit. The mass of the $B_c^{\pm}$ 
is taken from~\cite{bcmass}.
To take advantage of events with better resolution, the $K$-factor is applied in the analysis in six 
bins of $M(J/\psi\mu)$.

An unbinned likelihood fit is used to measure the average lifetime,
maximizing $\mathcal{L}$ over all $i$ candidates, where
\begin{eqnarray}
\mathcal{L}=\prod_{i}[f_{JT}\mathcal{F}_{JT}^i+(1-f_{JT})(f_{CB}\mathcal{F}_{CB}^i +\nonumber\\
(1-f_{CB})\{f_{SI}\mathcal{F}_{SI}^i+f_{JM}\mathcal{F}_{JM}^i+f_{B^+}\mathcal{F}_{B^+}^i+\nonumber\\
(1-f_{SI}-f_{JM}-f_{B^+})\mathcal{F}_{PR}^i\})].
\label{barebonesloglike}
\end{eqnarray}

Each component $\mathcal{F}$ consists of a 
combination of a mass-shape template
and a lifetime functional model, each described below,
to allow for a simultaneous fit of the fraction components and $\tau(B_c^{\pm})$. The fractions $f$ of each component have been described earlier. 
The fraction $f_{JT} = 0.034 \pm 0.002$ is taken from fits to the $B^+$ peak and 
$f_{CB} = 0.667 \pm 0.004$, is found
from \jpsi mass sideband fits. 
The exponential function $\mathcal{F}_{SI}$ is
convoluted with a Gaussian resolution function and 
smeared with normalized $K$-factor distributions. 
The width of the Gaussian resolution function 
uses the event-by-event uncertainty $\sigma(\lambda_i)$ on the VPDL, 
multiplied by a floating scale factor $s$ to take into account
any systematic underestimate of $\sigma(\lambda_i)$ due to tracking systematic 
uncertainties. 
A double Gaussian function, centered at VPDL = 0, is used
to model $\mathcal{F}_{PR}$.
The width of the inner Gaussian is given by $s \cdot \sigma(\lambda_i)$, and
the multiplicative factor for the width of the outer Gaussian is
determined using MC samples and data candidates with negative decay length.
Fits are made to the respective VPDL distributions to obtain $\mathcal{F}_{JT}$
 and $\mathcal{F}_{SB}$.
Empirical functional forms are used for both as fixed shapes, and
 the normalizations via the fractions $f_{JT}$ and
$f_{SB}$ are allowed to float in the fit. 
$\mathcal{F}_{JM}$ consists of a negative-slope exponential and
two positive-slope exponentials.  Starting values for the functional
parameters are determined from fits to the inclusive $J/\psi$ MC sample.
The normalization of the negative-slope exponential, along with
the normalization and slope of one of the positive-slope exponentials
are constrained by these fits.  The slope of the negative-slope and second
positive-slope exponential are allowed to float freely in the final fit.
%
$\mathcal{F}_{B^+}$ is a single exponential function with slope
constrained to the world-average value~\cite{PDG} convolved
with the same Gaussian resolution function as for the signal
lifetime model.

Before examining the fit to the data, possible lifetime biases are studied. 
Signal MC samples mixed with background are generated with 
different lifetimes.  Fits to these samples and ensemble tests indicate no significant bias and demonstrate the validity of the extracted statistical uncertainty. 

\begin{figure}[htb]\begin{center}
\centerline{\includegraphics[width=7cm]{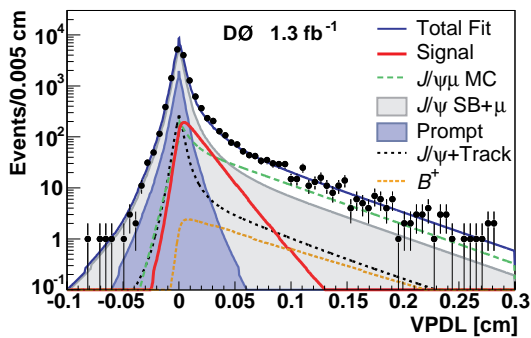}}
\caption{VPDL distribution of the $J/\psi \mu$ sample with the projected components of the fit overlaid. (color online)}
\label{finalfitdisplay}
\end{center}
\end{figure}

A simultaneous fit to the invariant mass and VPDL distributions
is performed using all the components described above. 
The fitted lifetime of the $B_c^{\pm}$ meson is found to be
$\tau(B_c^{\pm}) = 0.448^{+0.038}_{-0.036} 
\thinspace {\mathrm{(stat)}},$
with an estimated signal sample of 
$881 \pm 80\thinspace {\mathrm{(stat)}}$ candidates.
The fitted value of the scale factor is $s = 1.35 \pm 0.02$.
Figure~\ref{finalfitdisplay}  
shows the VPDL distribution of the $J/\psi$+$\mu$ sample with projections of the fit result overlaid. 

Stability checks made by dividing the data in half based on various selections show
no significant lifetime variations.
%
The systematic uncertainties considered are discussed in detail below and are summarized in
Table~\ref{systable}.

Variations of the $B_c^{\pm}$ mass within its measurement uncertainties~\cite{bcmass} make a negligible
difference in the lifetime.
The $B_c^{\pm}$ signal modeling uncertainty is estimated from the difference between the default and phase space decay models. 
The uncertainty in $p_T(B_c^{\pm})$ is found by reweighting the spectrum to correspond to varying the factorization and renormalization scales $\mu_F = \mu_R = \sqrt{p_T^2({\mathrm{parton}}) + m_b^2}$ by factors of a half and two~\cite{bc_prod}.
To address uncertainties on the predicted $p_T(b)$ for the signal and background component distributions, a momentum weighting factor that is applied to MC samples to improve the simulation and include the effects of the triggers is removed.
In all of the above cases, both
new signal mass templates and $K$-factor distributions are generated and the analysis repeated.
To assess the systematic uncertainty due to the modeling of the inclusive
$J/\psi$ MC 
mass distribution, 
contributions due to $b\bar{b}$ production via gluon splitting and
then flavor excitation are entirely removed. The shape of the mass template for the prompt component is varied within the statistical errors of its determination, and
the observed lifetime variation assigned as a systematic error. All of the systematic uncertainties described above are added in quadrature and summarized in Table 1 under the category of mass model uncertainties.

To test the assumption that the modeling of the lifetime of the
$J/\psi$ combinatoric background
can be approximated by
taking the average of the upper and lower mass sidebands,
the fit is performed using only the high or the low mass sideband, and a systematic uncertainty of one half the resulting shifts in lifetime is
assigned.
The scale factor $s$ is varied over the range of values, $1.2 - 1.4$, observed in other
lifetime analyses~\cite{sfact} as well as assigned a functional form and the variation in lifetime assigned as a systematic
uncertainty.
In the modeling of the prompt lifetime PDF, 
a single Gaussian function rather than a double Gaussian is used to 
describe the zero lifetime events. 
The shape parameters of the sideband lifetime model are changed by
varying the fit parameters within their uncertainties.  
The parameters defining the \jpsi QCD MC lifetime model  
are varied around their central values by $\pm$1$\sigma$.
For the $B^+$ lifetime model, 
the central value is changed by $\pm$1$\sigma$.
The $B^+$ lifetime is also allowed to float as a systematic study on the $B_c^{\pm}$ lifetime as well as a check of the $B^+$ lifetime, finding a value of $1.88 \pm 0.19$~ps, consistent with the 
world average value of $1.638 \pm 0.011$~ps~\cite{PDG}. 
All of the systematic uncertainties described above are added in quadrature under the category of lifetime model uncertainties.

Smaller systematic uncertainties arise from the variation of the fraction of the feed-down $B_c \rightarrow \psi(2S) X$ signal component between 0\% and 13\%, and from possible alignment  effects, estimated using signal MC with a modified detector geometry within the alignment tolerances.

\begin{table}[h]
\begin{center}
\caption{Summary of estimated systematic uncertainties.}
\begin{tabular}{lr}
\hline
\hline
Systematic  source& $\Delta \tau$~(ps) \\
\hline     
Mass model uncertainty& $\pm$0.021\\
Lifetime model uncertainty & $\pm$0.022\\
Signal feed-down fraction &  $\pm$0.005 \\
Alignment & $\pm$0.006\\

\hline
Total & $\pm$0.032\\
\hline
\hline
\label{systable}
\end{tabular}
\end{center}
\end{table}

In summary,
using approximately 1.3 fb$^{-1}$ of data, 
the lifetime of the $B_c^{\pm}$ meson is measured in the \bcdec final states. Using an unbinned likelihood simultaneous fit to the $J/\psi$+$\mu$ 
invariant mass and lifetime distributions we measure
$$\tau(B_c^{\pm}) = 0.448^{+0.038}_{-0.036} 
\thinspace {\mathrm{(stat)}} 
\pm 0.032 \thinspace {\mathrm{(sys)}}\thinspace {\mathrm{ps}}.
$$


This measurement is consistent with theoretical predictions of $0.55 \pm 0.15$~ps in an operator product expansion calculation~\cite{bc_pred} and 
$0.48 \pm 0.05$~ps using QCD sum rules~\cite{bc_pred2}. It is also consistent with the most recent measurement from the CDF collaboration~\cite{cdfbclife},
but with
significantly better precision, making this  measurement
of $\tau(B_c^{\pm})$ the most precise to date.

%
We thank the staffs at Fermilab and collaborating institutions, 
and acknowledge support from the 
DOE and NSF (USA);
CEA and CNRS/IN2P3 (France);
FASI, Rosatom and RFBR (Russia);
CNPq, FAPERJ, FAPESP and FUNDUNESP (Brazil);
DAE and DST (India);
Colciencias (Colombia);
CONACyT (Mexico);
KRF and KOSEF (Korea);
CONICET and UBACyT (Argentina);
FOM (The Netherlands);
STFC (United Kingdom);
MSMT and GACR (Czech Republic);
CRC Program, CFI, NSERC and WestGrid Project (Canada);
BMBF and DFG (Germany);
SFI (Ireland);
The Swedish Research Council (Sweden);
CAS and CNSF (China);
and the
Alexander von Humboldt Foundation.
%

\end{document}